\documentclass[a4paper,fleqn,usenatbib]{mnras}

\usepackage{newtxtext,newtxmath}

\usepackage[T1]{fontenc}
\usepackage{ae,aecompl}
\usepackage{booktabs}

\usepackage{graphicx}	
\usepackage{amsmath}	
\usepackage{amssymb}	
\usepackage{color}

\usepackage{float}

\usepackage[normalem]{ulem}

\newcommand{\iso}[2]{\hbox{${}^{#1}{\rm #2}$}}
\usepackage{adjustbox}
\usepackage{pifont}
\usepackage{float}
\usepackage[
  separate-uncertainty = true,
  multi-part-units = repeat
]{siunitx}
\usepackage{etoolbox}
\makeatletter
\patchcmd\@combinedblfloats{\box\@outputbox}{\unvbox\@outputbox}{}{\errmessage{\noexpand patch failed}}
\makeatother
\makeatletter

\newcommand{\Rmnum}[1]{\expandafter\@slowromancap\romannumeral #1@}
\makeatother

\title[S-process stars in LAMOST]{Discovery of s-process enhanced stars in the LAMOST survey}

\author[Brodie ~J. Norfolk et al.]{Brodie ~J. Norfolk,$^{1}$\thanks{E-mail: bnorfolk@swin.edu.au}\thanks{This paper includes data gathered with the 6.5 meter Magellan Telescopes located at Las Campanas Observatory, Chile.}
Andrew ~R. Casey$^{2,3}$,
Amanda ~I. Karakas$^{2}$,
Matthew ~T. Miles$^{2}$,\newauthor
Alex ~J. Kemp$^{2}$,
Kevin ~C. Schlaufman$^{5}$,
Melissa Ness$^{6}$,
Anna Y.~Q. Ho$^{4}$, \newauthor
John ~C. Lattanzio$^{2}$, 
Alexander ~P. Ji$^{7,8}$
\\
$^{1}$Centre for Astrophysics and Supercomputing (CAS), Swinburne University of Technology, Hawthorn, Victoria 3122, Australia\\
$^{2}$Monash Centre for Astrophysics (MoCA) and School of Physics and Astronomy, Monash University, Clayton Vic 3800, Australia\\
$^{3}$Faculty of Information Technology, Monash University, Clayton 3800, Victoria, Australia\\
$^{4}$Cahill Center for Astrophysics, California Institute of Technology, MC 249-17, 1200 E California Blvd, Pasadena, Ca, 91125, USA\\
$^{5}$Department of Physics and Astronomy, Johns Hopkins University, 3400 N Charles St., Baltimore, MD 21218, USA
\\
$^{6}$Department of Astronomy, Columbia University, 550 West 120th Street New York, New York 10027
\\
$^{7}$The Observatories of the Carnegie Institution for Science, 813 Santa Barbara St., Pasadena, CA 91101, USA \\
$^{8}$Hubble Fellow
}
\date{Accepted XXX. Received YYY; in original form ZZZ}

\pubyear{2018}

\begin{document}
\label{firstpage}
\pagerange{\pageref{firstpage}--\pageref{lastpage}}
\maketitle

\begin{abstract}

Here we present the discovery of 895 s-process-rich candidates from 454,180 giant stars observed by LAMOST using a data-driven approach. This sample constitutes the largest number of s-process enhanced stars ever discovered. Our sample includes: 187 s-process-rich candidates that are enhanced in both barium and strontium, 49 stars with significant barium enhancement only, and 659 stars that show only a strontium enhancement. Most of our sample are in the range of effective temperature and $\log{g}$ typical of RGB populations, which is consistent with our observational selection bias towards finding RGB stars. We estimate only a small fraction ($\sim$0.5\%) of binary configurations are favourable for s-process enriched stars. The majority of our s-process-rich candidates ($95\%$) show strong carbon enhancements however, only 5 candidates ($<3$\,\%) show evidence of sodium enhancement. Our kinematic analysis reveals our sample is composed of 97\% disk stars, with the other 3\% showing velocities consistent with the Galactic halo. The scale height of the disk is estimated to be $z_h=0.634 \pm{0.063} kpc$, comparable to literature values. A comparison with yields from asymptotic giant branch (AGB) models suggests the main neutron source responsible for the Ba and Sr enhancements is the \iso{13}C($\alpha$,n)\iso{16}O reaction. We conclude that s-process-rich candidates may have received their over-abundances via mass-transfer from a previous AGB companion with an initial mass in the range $1 - 3\,M_{\odot}$. 
\end{abstract}


\section{Introduction} \label{sec:intro}

The abundance of elements created through the slow neutron capture process (s-process) is a measure of the nucleosynthetic reactions that occur primarily in Asymptotic Giant Branch (AGB) stars. Due to their associated enrichment of the s-process element barium, stars with peculiar enhancements of carbon and heavy elements (Z > 30) are commonly referred to as `barium stars' \citep{Bidelman1951}. The naming convention associated with {\em barium stars} is historical, and for the purposes of this paper it is necessary to clarify the definitions that will be used in subsequent discussions. Throughout this work we will refer to stars with Ba and Sr enrichment as s-process-rich candidates, and stars with only strong Sr enrichment or only strong Ba enrichment as Sr-rich or Ba-rich candidates respectively. Metal-poor stars ([Fe/H] $<-1.2$) enriched by s-process elements are typically referred to as CH stars in the literature; here we will simply describe our discoveries within this metallicity range as metal-poor s-process-rich candidates.

Extrinsic or intrinsic mechanisms can be invoked to explain s-process enrichment in a given star. This refers to the source of the enrichment: either a mechanism intrinsic to the star (eg. an internal nuclear process), or an extrinsic process (e.g. pollution by stellar winds from a binary companion) that results in an over-abundance of s-process elements. Both intrinsic and extrinsic mechanisms for enhancement may contribute somewhat to the known population of s-process enhanced stars.

Stars intrinsically enhanced in s-process elements are in the thermally-pulsing AGB phase of stellar evolution. During this stage of evolution, nucleosynthesis occurring in the helium shell followed by subsequent mixing can result in an overabundance of carbon and s-process elements \citep{herwig2005,karakas2014} at the surface.
Extrinsic enhancement occurs according to the mass-transfer hypothesis. According to the mass-transfer hypothesis, extrinsic s-process stars can be formed by stellar wind accretion \citep{boffin1988,jorissen1992,han1995} or Roche-lobe overflow \citep{webbink1986,han1995} onto a lower mass main sequence companion. The mass-transfer hypothesis assumes that the currently observed s-rich star is in a binary system and received the overabundance of heavy elements from an AGB companion which has since evolved into a white dwarf. In fact, radial-velocity observations of s-rich stars show long-term velocity variations \citep{mcclure1983}, suggesting that all barium stars are in binary systems.
For these reasons, the properties and occurrence rate of s-process enrichments in RGB populations is informative of the binary star fraction as a function of metallicity, the mass ratio of binary stars, as well as how AGB yields vary across a wide range of initial masses and metallicities.

\begin{figure*}
	\includegraphics[width=\textwidth]{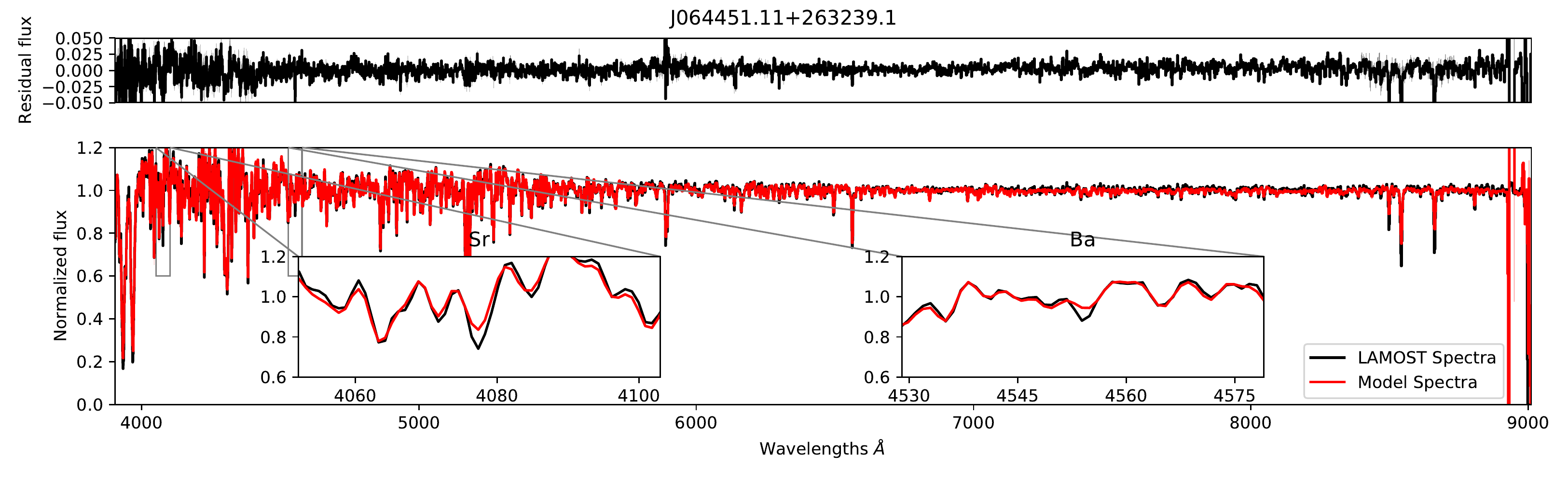}
	\caption{Pseudo-continuum-normalised LAMOST spectra for the s-process candidate J064451.11+263239.2. The data are shown in black and the best-fitting data-driven model is shown in red. We include zoom-in axes to show significant deviations in Sr and Ba at  4077\,\AA\ and 4554\,\AA\, respectively.}
	\label{fig:figure1}
\end{figure*}

The s-process synthesizes roughly half of all elements heavier than iron \citep[e.g.,][]{busso1999,travaglio2001,herwig2005,bisterzo2014,karakas2014}. During the thermally-pulsing AGB phase, thermal instabilities occur in the He shell every $10^5$ years or so, depending on the mass of the H-exhausted core. These bursts of energy drive a convective zone across almost the entire region lying between the base of the He-shell and H-shell. This mixes the products of nucleosynthesis within these regions while causing a radial expansion pushing the H-shell out to cooler regions \citep{karakas2002}. Following a thermal pulse, the convective envelope can then move inwards towards regions previously mixed by the thermal pulse driven convective zones. This inward movement of the convective envelope is known as the third dredge up (TDU), and may occur after each thermal pulse. During the TP-AGB phase, the TDU is responsible for the surface enrichment in \iso{12}C and heavy elements produced by the s-process \citep[e.g.,][]{busso2001}. Following the third dredge up the star contracts, reigniting the H-shell and preventing further mixing to the surface. The interpulse, thermal pulse, and dredge up cycle may occur many times.

The progenitors of s-process-rich stars evolve from a metallicity-dependent initial mass range of approximately $0.8 - 8\,M_{\odot}$ \citep{karakas_lugaro2016}. The minimum mass ($0.8\,M_{\odot}$) of these progenitor AGB stars is defined by the onset of core helium burning while the maximum ($8\,M_{\odot}$) is determined by the onset of core carbon burning. The age of these progenitor stars varies considerably, with lower mass stars reaching ages $\approx$ 12\,Gyr in metal-poor globular clusters. More massive metal-rich stars may have ages of less than $\approx$ 100Myr, including progenitor stars that are at the core carbon burning limit or very close to it \citep[e.g.,][]{whitelock2013}.

The majority of s-process-rich stars are observed to be members of the disk. \citet{gomez1997} finds 90\% percent of their 318 barium stars have velocities characteristic of the disk, \citet{pereira2011}'s sample of 12 barium stars are fully consistent with disk membership, and 90\% of \citet{decastro2016}'s 182 barium stars populate the Galactic disk. \cite{jorissen1993} relates galactic position to the extrinsic or intrinsic nature of their s-process-rich stars. They show intrinsic s-process-rich stars are concentrated towards the galactic plane, whereas extrinsic s-process-rich stars are uniformly distributed in absolute galactic latitude.

In this paper we analyse 454,180 giant stars from the second LAMOST data release \citep{luo2015}, and identify 895 s-process-rich candidates using a data-driven approach. This involved filtering for significant flux residuals at the Sr and Ba lines. In Section \ref{sec:methods} we describe our observations and the candidate selection process. Section \ref{sec:observations} details the analysis of high-resolution follow-up observations obtained for a few candidates. In Section \ref{sec:dis} we discuss the properties of our s-process-rich candidates in context of existing literature. We provide concluding remarks in Section \ref{sec:con}.

\section{Methods} \label{sec:methods}
\subsection{LAMOST analysis}
\subsubsection{Data-driven analysis}
The LAMOST (Large sky Area Multi-Object Fibre Spectrographic Telescope) survey released low-resolution ($\mathcal{R} \approx 1800$) optical spectra (3700\,\AA\ to 9000\,\AA) for 2,207,189 stars in their second data release \citep{luo2015}. For details regarding the LAMOST atmospheric parameters and error analysis we direct the reader to the LAMOST Stellar Parameter pipeline \citep{wu2011a,wu2011b,luo2015}. Stellar parameters and abundances ($T_{\rm eff}$, $\log{g}$, [M/H], [$\alpha$/M]) were derived by \citet{ho2017} using \emph{The Cannon} \citep{ness2016}. This involved label transfer (a process for improving the accuracy of parameters and abundances in one survey using results from another) from APOGEE to LAMOST. This was accomplished using 9,952 giant stars in common as a training sample to generate a predictive model which was then applied to the stars that make up our 454180 star sample. As a consequence of this procedure, our sample is limited to giant stars, and we are therefore unable to identify dwarf s-process enhanced stars. Cross-validation tests between low-resolution ($\mathcal{R} \approx 1800$) LAMOST spectra and higher-resolution ($\mathcal{R} \approx 22,500$) APOGEE spectra substantially reduce the inconsistencies and label uncertainties between each individual survey. The typical uncertainties are approximately 70\,K in effective temperature $\rm T_{\rm eff}$, 0.1\,dex in surface gravity $\log{g}$, 0.1\,dex in metallicity [M/H], and 0.04\,dex in the abundance of $\alpha$-elements relative to overall metallicity [$\alpha$/M]. These uncertainties are comparable to the those present in APOGEE \citep{alam2015}.

\subsubsection{Candidate selection} \label{sec:cand}
The s-process-rich candidates in this work were identified by filtering for significant flux residuals at the Sr and Ba lines. We categorize each spectrum as s-process enriched, Ba-enriched, or Sr-enriched. We fit a Gaussian function with amplitude $A$ to the residuals of the observed flux and the data-driven model from \emph{The Cannon}, and the standard deviation $\sigma_A$ at wavelengths 4554\,\AA\ (\ion{Ba}{II}) and 4077\,\AA\ (\ion{Sr}{II}) for s-process-rich candidates, at wavelengths 4554\,\AA\ (\ion{Ba}{II}) and 4934\,\AA\ (\ion{Ba}{II}) for barium-enriched candidates, and at wavelengths 4077\,\AA\ (\ion{Sr}{II}) and 4215\,\AA\ (\ion{Sr}{II}) for strontium enriched candidates. The amplitude of each Gaussian fit at each wavelength is a measure of the depth of each absorption line relative to what is expected of a star of that metallicity, alpha-abundance, effective temperature, and $\log{g}$. If the observed spectrum has deeper absorption lines than the model at the corresponding wavelengths, it is considered a candidate with an enhancement. A randomly chosen s-process candidate is shown in Figure \ref{fig:figure1}.

We used five criteria for each spectrum and subsequent set of absorption lines in order to identify potential s-process enrichment:

\renewcommand\labelenumi{(\roman{enumi})}
\renewcommand\theenumi\labelenumi

\begin{enumerate} 
\item Profile amplitude $A$ for both enhanced lines must be $\Delta A < -0.05$, indicating a stronger absorption line than expected by the model.
\item Both amplitudes must be measured within 3 standard deviations of the profile amplitude ($|\Delta A|/\sigma _A$ > 3).
\item The wavelength at each (Doppler-corrected) absorption line must be within 2\,\AA\ ($\Delta \lambda$ < 2 \AA) of the rest frame laboratory wavelength.
\item The reduced $\chi^2$ from \emph{The Cannon} must satisfy $\chi_r^2 < 3$.
\item The LAMOST spectra must have a signal-to-noise ratio of $S/N > 30\,\textrm{pixel}^{-1}$.
\end{enumerate}
In addition to requiring matches at two wavelengths (e.g., both 4934\,\AA\ and 4554\,\AA\ for barium-enriched candidates), we visually inspected all candidates to exclude spectra with data reduction issues, apparent absorption finer than the spectral resolution, or overly noisy spectra. This approach yielded 895 candidates, the details of which are listed in Table 1, including 187 s-process-rich candidates, 49 barium-enriched stars, and 659 stars with just strontium enrichment. This constitutes the largest collection of s-process enhanced stars ever discovered. The previous most numerous sample is the 182 barium-enriched (\ion{Ba}{II}) stars presented by \citet{macconnell1972}.

For the remainder of this paper we restrict our analysis to the 187 s-process-rich candidates that show enhancement in both barium and strontium. While the 49 or 659 stars with only barium or strontium enhancements (respectively) may also be classified as s-process-rich candidates, there may be other explanations for their chemical abundance pattern \citep[e.g.,][]{maiorca2011}. 

\subsubsection{Enhancements due to sodium, technetium, and carbon} \label{sec:other enhancements}
Enhancements in sodium, technetium, and carbon are useful for determining whether a sample is populated by AGB stars or by polluted extrinsic s-process-rich stars. We performed an identical analysis to the process described in Section \ref{sec:cand} in order to identify negative flux residuals for enhancement in sodium, technetium, or carbon. For sodium enrichment, we required significant absorption in the doublet lines at 5889\,\AA\ and 5895\,\AA. Only 5/187 s-process-rich candidates met these criteria. For technetium enhancement, we searched for significant residual deviations at 4049\,\AA, 4238\,\AA, 4262\,\AA, 4297\,\AA, and 5924\,\AA. These absorption lines are extremely weak and require a substantial amount of technetium enhancement before it would become visible even in a high S/N LAMOST spectrum. We found that 51 s-process-rich candidates exhibited some level of significant enhancement at the single absorption line 4238\,\AA. However we caution that Tc enhancement is usually a very weak signature in such warm stars and would require multiple lines of enhancement, and is considerably blended by other s-process element lines \citep[e.g.,][]{van1999}. As such, we discard these 51 stars with enhancement at a single Tc absorption line as false positives. For carbon enhancement we searched for significant deviations at the G band of CH near 4300\,\AA, of which 178/187 s-process-rich candidates exhibited significant flux residuals and are therefore considered carbon enriched.

\begin{table*}
	\centering
	\caption{Properties of 895 s-process-rich candidates. The table is available online in its entirety, here we show a portion to demonstrate its style and content.}
	\label{table:table1}
	\begin{adjustbox}{width=1\textwidth}
		\begin{tabular}{@{}|l|l|c|c|c|c|c|c|c|c|c|c|c|c|c|@{}}
			\toprule
			2MASSID             & R.~A.         & Dec.        & $v_{r}$ & S/N & $\rm T_{\rm eff}$ & $\log{g}$ & [Fe/H] & [$\alpha$/Fe] & $\chi_r^2$ & [Ba/Fe] & [Sr/Fe] & \ion{Ba}{II} & \ion{Sr}{II} &  \ion{Ba}{II} \& \ion{Sr}{II} \\
			& (J2000) & (J2000) & (km\,s$^{-1}$) & (pixel$^{-1}$) & (K)  \\ \midrule
			J000019.26+501444.8 & 00:00:19.27 & +50:14:44.9 & -3.9  & 49      & 4973         & 3.27         & 0.21         & 0.08             & 0.66               & 0.25        & 0.83        & \ding{55} & \ding{51}  & \ding{55}   \\
			J000020.55+411348.1 & 00:00:20.56 & +41:13:48.2 & -28.2 & 32      & 4882         & 2.74         & -0.22        & 0.04             & 0.23               & -0.17       & 0.90        & \ding{55}& \ding{51}  & \ding{55}    \\
			J000134.95+490743.2 & 00:01:34.96 & +49:07:43.2 & -42.3 & 72      & 5044         & 3.11         & -0.54        & 0.11             & 0.79               & 1.02        & 0.45        & \ding{55} & \ding{55} & \ding{51}  \\
			J000258.09+410730.0 & 00:02:58.10 & +41:07:30.1 & -34.5 & 41      & 4697         & 2.57         & -0.22        & 0.12             & 0.98               & -0.10       & 0.80        & \ding{55} & \ding{51}  & \ding{55}  \\
			J000403.80+160257.1 & 00:04:03.80 & +16:02:57.2 & -35.4 & 42      & 5200         & 3.40         & -0.41        & 0.09             & 0.33               & 0.92        & 0.52        & \ding{55} & \ding{55} & \ding{51}  \\
			J000439.16+183350.3 & 00:04:39.17 & +18:33:50.4 & -38.7 & 31      & 4601         & 2.50         & 0.44         & 0.03             & 0.34               & -0.11       & 0.88        & \ding{55} & \ding{51}  & \ding{55}  \\
			J000444.87+400402.1 & 00:04:44.88 & +40:04:02.1 & -94.7 & 53      & 4172         & 1.46         & -0.08        & 0.09             & 0.72               & 0.03        & 0.91        & \ding{55} & \ding{51}  & \ding{55}  \\
			J000552.76+261849.3 & 00:05:52.76 & +26:18:49.4 & -22.5 & 73      & 4924         & 3.14         & -0.05        & 0.09             & 0.67               & 0.28        & 0.85        & \ding{55} & \ding{51}  & \ding{55}  \\
			J000737.70+394055.5 & 00:07:37.70 & +39:40:55.6 & -31.5 & 33      & 4700         & 2.80         & -0.04        & 0.07             & 0.28               & -0.13       & 0.81        & \ding{55} & \ding{51}  & \ding{55}  \\ \hline
		\end{tabular}
	\end{adjustbox}
\end{table*}

\subsubsection{Abundances estimated from LAMOST spectra}
We estimated [Ba/Fe] and [Sr/Fe] abundance ratios for all s-process-rich candidates by spectrum synthesis. We assumed that absorption due to most metal lines is captured by \emph{The Cannon} model, and deviations in flux at the 4554\,\AA\ \ion{Ba}{II} line and the 4077\,\AA\ \ion{Sr}{II} transition are solely due to enhancements in Ba and Sr, respectively. We used the grid of MARCS model atmospheres \citep{marcs}, atomic transitions from VALD \citep{vald}, and the SME code \citep{sme,ispec} to synthesize spectra of each star, then varied the Ba and Sr abundances until they matched the flux deviations away from our data-driven model. We adopted the stellar parameters ($T_{\rm eff}$, $\log{g}$, [Fe/H]) from \citet{ho2017}, and assume a microturbulence velocity of $v_{mic} = 2\,{\rm km\,s}^{-1}$. Uncertainties in [Ba/Fe] and [Sr/Fe] from these spectra are taken as the fitting error due to noise, added in quadrature with an adopted $0.2\,{\rm dex}$ systematic error floor. We assume the same isotopic fractions from \citet{sneden2008}. Previous studies have used a threshold of $\textrm{[Ba/Fe]} > +0.3$ to define an s-process-rich star \citep{malaney1988}. Our abundance analysis shows 186/187 of our s-process-rich candidates -- identified from flux residuals alone -- meet this criterion based on spectrum synthesis of the 4554\,\AA\ transition. Table \ref{table:table1} lists a portion of our 895 s-process enhanced sample.

\subsection{Dynamics}
We integrated the galactic orbits using astrometry from the second Gaia data release \citep{gaia2016,gaia2018b, cropper2018, katz2018, lindegren2018, sartoretti2018} for 871/895 of the s-process enhanced stars with positive parallaxes ($\varpi > 0$) and parallax S/N of five or above ($\varpi/\sigma_\varpi > 5$). We integrated each star backwards for $0.5\,\textrm{Gyr}$ using the \texttt{gala} Python package \citep{price2017} in a Milky Way-like potential \citep{mwpotential2014} that consists of three components: a \citet{hernquist1990} bulge and nucleus, a \citet{miyamoto1975} disk, and a \citet{nfw1997} halo. We computed spatial velocities relative to the local standard of rest, where $U_{LSR}$ is positive towards the Galactic centre, $V_{LSR}$ is positive in the direction of Galactic rotation ($l=90^{\circ}, b=0^{\circ}$), and $W_{LSR}$ is positive towards the north galactic pole ($b=90^{\circ}$). Our analysis revealed 97\% disk star membership and 3\% halo star membership.

\section{Follow-up observations with Magellan/MIKE} \label{sec:observations}

\subsection{Observations and data reduction}
High-resolution spectra were obtained for two of the s-process-rich candidates (J09162834+0259348 and J08351472-0548480) using the MIKE (Magellan Inamori Kyocera Echelle) \citep{bernstein2003} spectrograph on the Magellan Clay telescope \citep{schectman2003} at Las Campanas Observatory, Chile. These two candidates were selected as targets of opportunity only, and do not represent a comprehensive follow-up observational campaign. We observed both stars in good seeing using the 0.7 arcsecond slit and 2x2 spatial on-chip binning, providing a spectral resolution of $\mathcal{R} \approx 28,000$. Exposure times of 100 seconds were sufficient to achieve a S/N ratio exceeding 30 per pixel at 4500\,\AA. We acquired calibration (biases, milky, quartz, and Th-Ar arc lamp) frames in the afternoon. We reduced the data using the \texttt{CarPy} package \citep{kelson2000}, used spline functions to continuum-normalise all echelle orders, and resampled the normalised spectra onto a uniform-spaced wavelength map. We used a rest-frame normalised template appropriate for a FGK-type star to place the observed spectra at rest.

\subsection{Abundance analysis}
We adopted the stellar parameters ($T_{\rm eff}$, $\log_{10}g$, [Fe/H]) provided from the data-driven analysis of \citet{ho2017}. Following the procedure outlined in \citet{casey2014}, we measured the strength of \ion{Ba}{II} (4554\,\AA, 4934\,\AA, and 6496\,\AA) and \ion{Sr}{II} (4077\,\AA\ and 4215\,\AA) absorption lines by spectral synthesis, using s-process isotopic ratios from \citet{sneden2008}. The abundance ratios we estimate from high-resolution spectra are in excellent agreement with our estimates from LAMOST spectra, all agreeing within the joint $1.2\sigma$.

For J09162834+0259348: from high- and low-resolution spectra, respectively, we find $[{\rm Sr/Fe}] = 0.76 \pm 0.10$ and $0.85 \pm 0.21$, and $[{\rm Ba/Fe}] = 0.92 \pm 0.10$ and $0.77 \pm 0.21$. The largest discrepancy we find between high- and low-resolution spectra is [Sr/Fe] for J08351472-0548480, where we find $[{\rm Sr/Fe}] = 0.62 \pm 0.07$ from our Magellan/MIKE spectra, compared with $0.90 \pm 0.22$ from LAMOST. For J08351472-0548480 we also find $[{\rm Ba/Fe}] = 0.94 \pm 0.12$ from high-resolution spectra and $[{\rm Ba/Fe}] = 0.80 \pm 0.26$ from low-resolution spectra. 

Uncertainties on abundances derived from high-resolution spectra are taken as the standard deviation of multiple line measurements. The high-resolution abundances we find help validate our methodology for candidate selection, and for estimating abundances from LAMOST spectra.


\section{Discussion}  \label{sec:dis}

\subsection{Extrinsic or intrinsic}
\begin{figure}
	\includegraphics[width=0.47\textwidth]{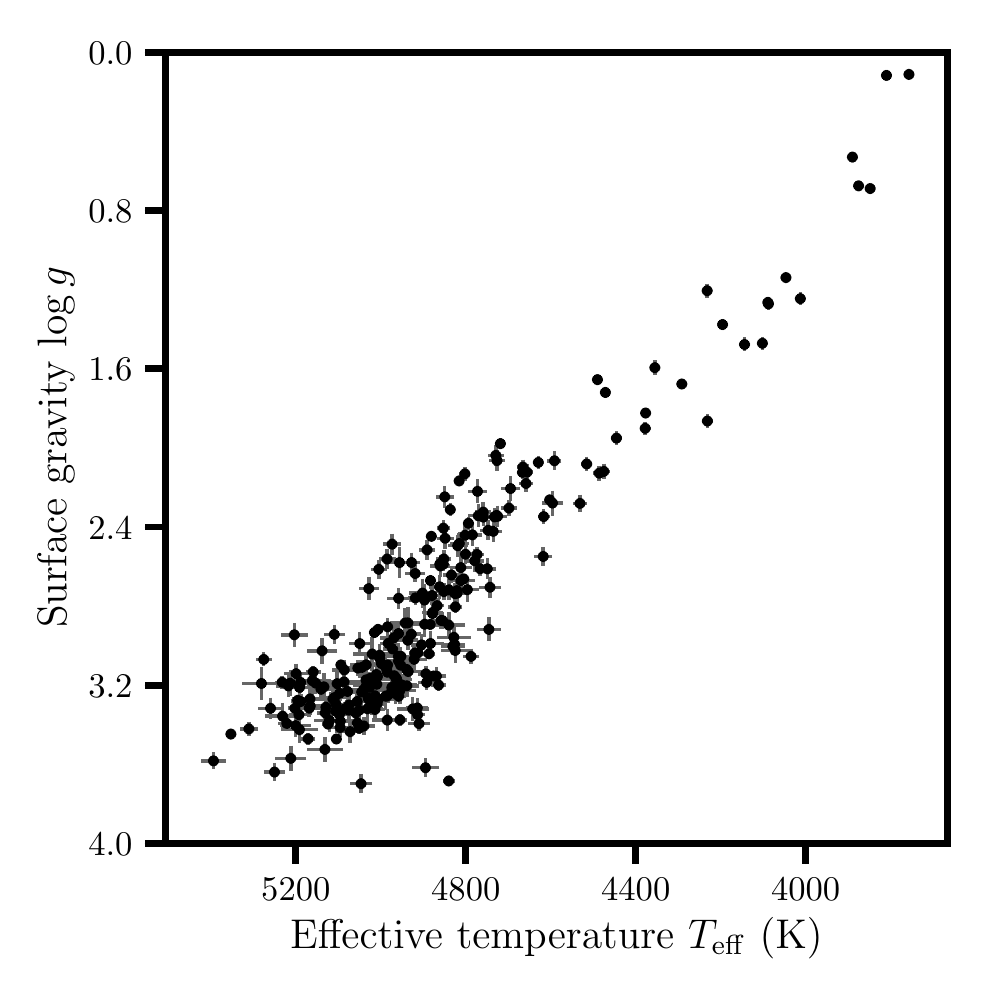}
    \caption{Effective temperature $T_{\rm eff}$ and surface gravity $\log{g}$ for 187 s-process-rich candidates from LAMOST.}
    
    \label{fig:figure2}
\end{figure}

S-process enrichment can be explained by intrinsic or extrinsic mechanisms. Intrinsically enhanced s-process-rich stars must be massive enough to reach the TP-AGB phase within the age of the Galaxy, and extrinsic s-process stars should be a member of a binary system where the companion is now a white dwarf. Evidence of a recent ($<1\,\textrm{Myr}$) mass pollution event can present itself as enrichment in both s-process elements and 99Tc. Figure \ref{fig:figure2} shows that the majority of our sample is within temperature and surface gravity ranges consistent with the first giant branch. This figure highlights that our sample does not appear to contain many stars evolved enough to have reached the TP-AGB phase and therefore, must be composed primarily of extrinsic s-process-rich candidates. Candidates with $\log{g} \lesssim 2$ in Figure \ref{fig:figure2} could be either AGB stars or RGB stars.


\subsection{Frequency of S-process Stars}
We find the frequency of s-process-rich stars among a well-defined sample of giants to be $\sim$0.2\%, substantially lower than the canonical value of 1\% from \citet{macconnell1972}. The majority of our sample appears to consist of extrinsic s-process-rich candidates polluted by a presumed binary companion. \citet{badenes2018} found the multiplicity inferred from APOGEE observational data is best explained by assuming a binary fraction of $\sim$35\%, and \citet{tian2018} estimate a binary fraction of $\sim$50\% for solar-type stars in LAMOST DR4. A fraction of those binary systems will become s-process enriched. However, we find only 0.2\% of systems in our sample are s-process-rich. Therefore taking the mean binary fraction estimate of 42.5\%, only $\sim$0.5\% of binary systems will end up as s-process enriched systems. This highlights that there is a set of constraining conditions that must be met in a binary system in order for it to result in an s-process-rich star \citep{jorissen2019}. Binary stellar evolution theory should be able to provide information on these constraints, however the population synthesis and physical modelling required to produce such information is beyond the scope of this work.

\subsection{Sodium enhancement}
Sodium is produced by the Ne-Na cycle in any hydrogen burning environment where the temperature exceeds about $30\times 10^{6}\,$K \citep{arnould1999}. This includes the convective cores of intermediate-mass stars and the H-burning shells of AGB stars \citep{el1995,mowlavi1999,karakas2003}. The surface of the giant becomes enriched in Na via the first dredge-up, although predicted enrichments are small (on the order of 0.15 dex for $M\lesssim 4\,M_{\odot}$), after second dredge-up where predicted enrichments are larger \citep{karakas2014}, and during the AGB.

Intermediate-mass AGB stars that experience hot bottom burning are predicted to produce Na, but model predictions depend on uncertain physics \citep[e.g.,][]{ventura2013,slemer2017}. It is unclear if mass-transfer from intermediate-mass AGB stars can produce a s-process candidate but it would likely be C and F poor, due to hot-bottom-burning, and rich in Sr over Ba \citep[e.g.,][]{karakas_lugaro2016} (See Sect \ref{sec:4.4} for further discussion). 



\citet{antipova2004} reported  enhanced sodium in 3 of their 16 RGB s-process-rich stars. They relate this to the first dredge-up of nuclear-burning material produced by convection during the red-giant phase, and suggest that [Na/Fe] ratios are systematically higher for giants with lower $\log{g}$ values. Similarly, \citet{decastro2016} propose a possible weak anti-correlation between [Na/Fe] ratio and $\log{g}$, and highlights that this trend is present in previous studies \citep[e.g.,][]{boyarchuk2002,mishenina2006,luck2007,takeda2008}.

\begin{figure}
	\includegraphics[width=0.47\textwidth]{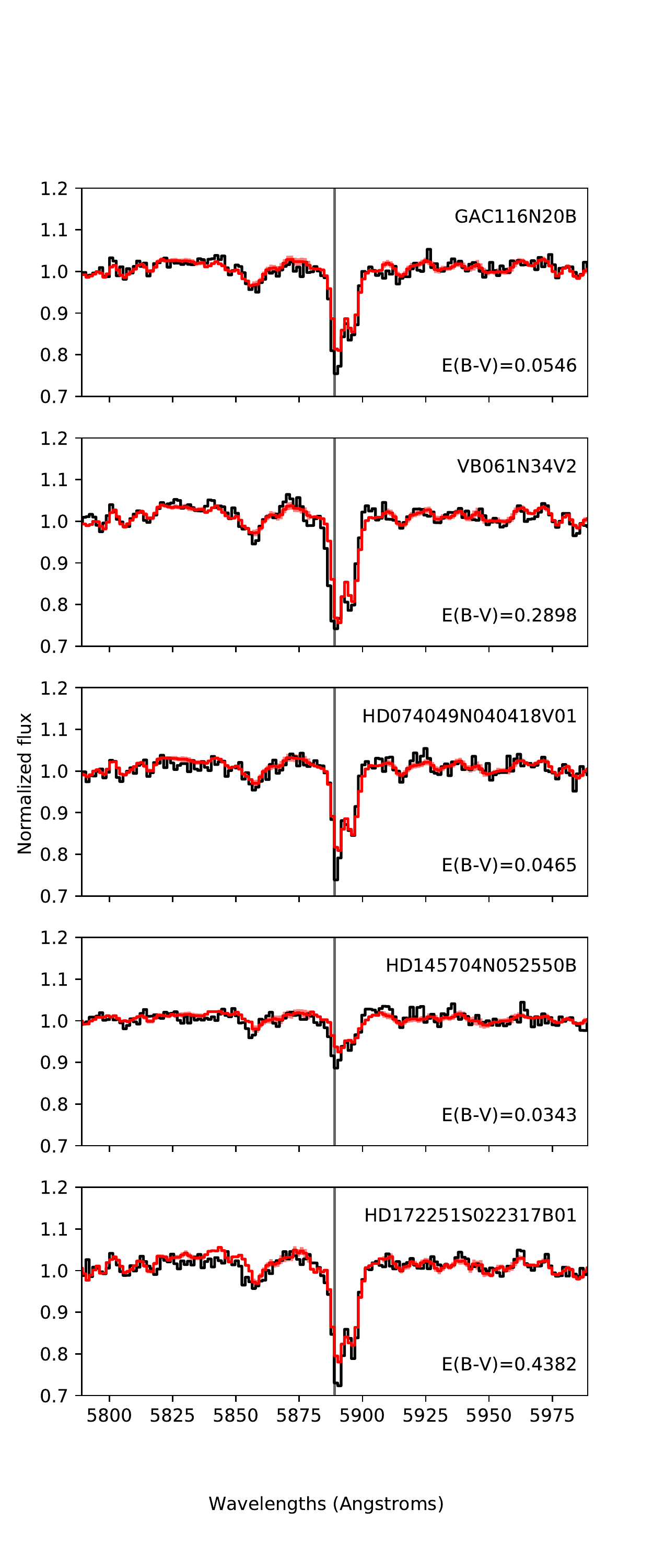}
    \caption{Pseudo-continuum-normalised LAMOST spectra for the s-process candidates enhanced in Na as shown by the deviations in the doublet lines at 5889\,\AA\ and 5895\,\AA. The data are shown in black and the best-fitting data-driven model is shown in red.}
    \label{fig:sodium}
\end{figure}

We find that only 5/187 s-process-rich candidates ($<3$\,\%) exhibit enhanced sodium as shown in Figure \ref{fig:sodium}, and all 5 have  $\log{g} \approx 3$. Moreover, it is likely that some of these stars appear to have high sodium due to interstellar gas lines of Na being misinterpreted as stellar absorption. 
Using the IRAS all-sky dust map \citep{schlafly2011}, we find that 2/5 of these candidates have E(B-V)$\approx$ 0.35, suggestive that the flux residuals around the Na doublet are due to interstellar absorption. However 3/5 candidates exhibit low ($\approx$0.045) E(B-V) values, indicating the enhancement is more likely due to stellar absorption. These results suggest that sodium-rich material from a companion TP-AGB star has polluted these 3 s-process-rich candidates. It also demonstrates that sodium enhancement is by no means ubiquitous in s-process-rich candidates. We do not observe a correlation between [Na/Fe] and $\log{g}$ in our sample. We also find that s-process-rich candidates do not have higher [Na/Fe] abundance ratios on average. 


\begin{figure}
	\includegraphics[width=0.47\textwidth]{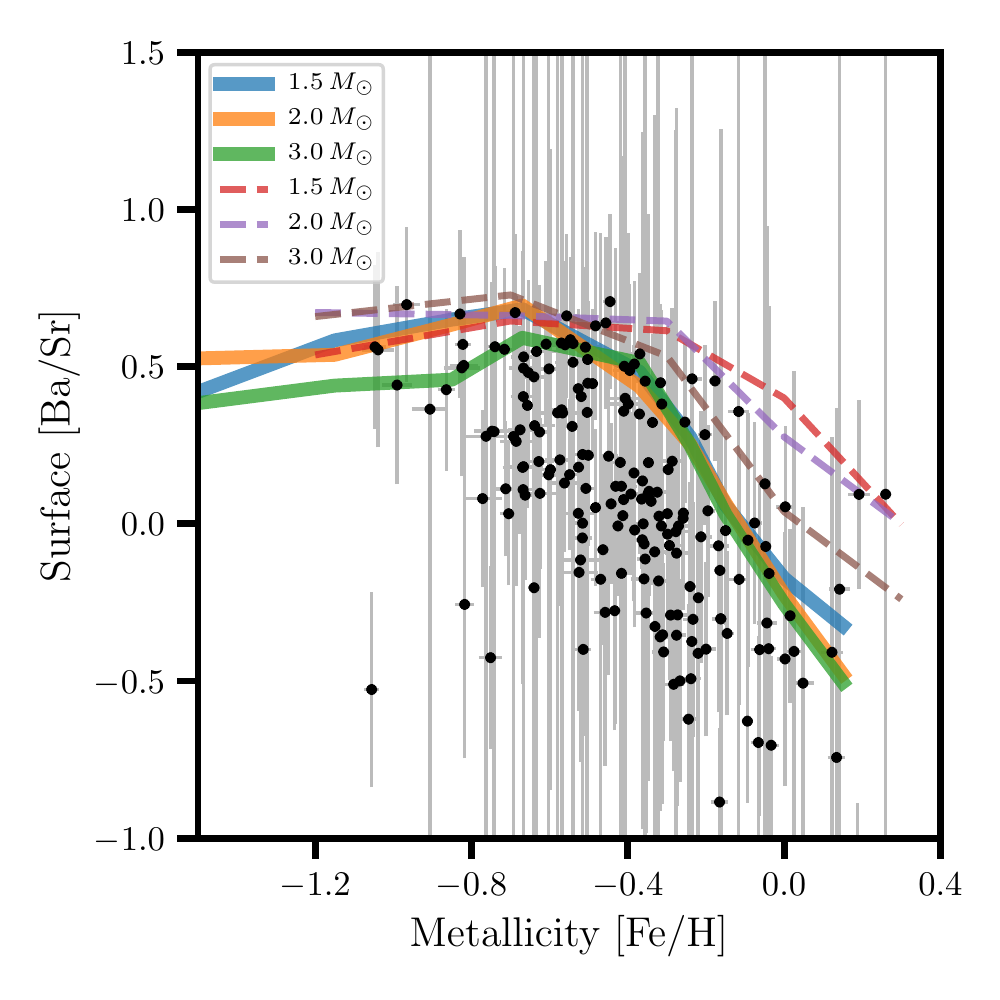}
    \caption{Metallicities ([Fe/H]; x-axis) and heavy-to-light s-process abundance ratios ([Ba/Sr] as measured from LAMOST; y-axis) for 187 s-process-rich candidates. Coloured lines blue, orange, and green, indicate surface [hs/ls] yields from \citet{cristallo2015} for different masses. Coloured lines red, purple, and brown indicate surface [hs/ls] yields from \citet{karakas_lugaro2016} for [Fe/H] at +0.3. 0.0, and -0.3; from \citet{karakas2018} for [Fe/H] $= -$0.7; and from \citet{fishlock2014} for [Fe/H] = $-$1.2.}
    \label{fig:figure3}
\end{figure}


\subsection{Comparison to AGB yields} \label{sec:4.4}
In Figure \ref{fig:figure3} we show the [Ba/Sr] abundance ratio (for all 187 s-process candidates identified in LAMOST). 
Although the [Ba/Sr] ratio is quite noisy ($\sim$0.6\,dex mean value of the uncertainty) the overall metallicity [Fe/H] is quite precise (0.1\,dex or better). The observed ratios of [Ba/Sr] shown in Figure \ref{fig:figure3} are particularly interesting for comparison to theoretical predictions of the $s$-process because we are comparing two elements that are only produced in AGB stars. \citet{lugaro2012} discussed how the ratios of light to heavy $s$-process elements are essentially independent of stellar modelling uncertainties including third dredge-up mixing, mass loss, and the accretion and mixing processes on the binary companion. Ratios between elements produced by neutron-capture are a measure of the thermodynamic conditions occurring in the previous AGB star, the neutron source, and the neutron flux. 

In Figure \ref{fig:figure3} we also show theoretical $s$-process predictions from two groups: The FRUITY\footnote{ http://fruity.oa-teramo.inaf.it/} yields from \citet{cristallo2015} which cover a range in metallicity from [Fe/H] $\approx 0.1$ to $-2.3$, and the Monash yields \citep{fishlock2014,karakas_lugaro2016} which cover a range in metallicity from [Fe/H] = $+0.3$ to $-1.2$ (as similarly done in \citet{cseh2018}). For both sets of models we show results from initial masses between 1.5$M_{\odot}$ and 3$M_{\odot}$.
Observational evidence suggests that the dominant polluters of s-process-rich candidates are low-mass AGB stars with masses below $\lesssim 4M_{\odot}$ where the main neutron source is the $\text{C}^{13}$(\textrm{$\alpha$},n)$\text{O}^{16}$ reaction \citep[e.g.,][]{lugaro2012, karinkuzhi2018}.

Reasons for this include the fact that intermediate-mass stars over approximately 4$M_{\odot}$ do not become carbon rich on the AGB (or if they do, only at the very end), and that intermediate-mass stars near solar metallicity are not predicted to produce much barium at all. This is partly a consequence of the neutron source operating inside intermediate-mass stars, that is the $\text{Ne}^{22}$(\textrm{$\alpha$},n)$\text{Mg}^{25}$ source which favours production of Sr over Ba \citep[e.g.,][]{karakas_lugaro2016}. It is unclear if the $\text{C}^{13}$(\textrm{$\alpha$},n)$\text{O}^{16}$ neutron source, which dominates in low-mass AGB stars, also operates in intermediate-mass AGB stars (see discussions in \citet{straniero2014}, \citet{garcia2013}). 

Stellar yields are the integrated mass expelled into the interstellar medium, and represent an upper limit that includes surface abundance changes from all previous mixing events. Therefore it is not surprising that the yield predictions are in the upper range of the observed ratios. Mass transfer from an AGB star to a companion does not necessarily occur right at the end of the AGB star's life. Mass transfer via a stellar wind can occur any time that the AGB star is losing mass via a sufficiently strong wind, and mass transfer via Roche-lobe overflow can occur at any time during the AGB phase, regardless of whether the star is experiencing significant wind mass-loss (see population synthesis studies of peculiar stars \citep[e.g.][]{han1995, karakas2000, liu2000, abate2013, abate2015}).

For example, if we examine the 2$M_{\odot}$, [Fe/H] = 0 model from \citet{karakas_lugaro2016} and calculate the [Ba/Sr] ratio as a function of third dredge-up mixing episodes, we see that the ratio increases from 0.05 at the first mixing event to 0.24 by the second before reaching a constant value of 0.35 by the 6th. Ratios of [Ba/Sr] $<0$ are not predicted from low-mass AGB models that become C-rich, which is a consequence of the $\text{C}^{13}$(\textrm{$\alpha$},n)$\text{O}^{16}$ neutron source causing high neutron exposures and element production at the Ba and Pb peaks \citep[e.g.][]{busso2001}. It is possible for an intermediate-mass AGB star to produce [Ba/Sr] $<0$, e.g., the 5$M_{\odot}$, [Fe/H] = $-0.7$ model from \citet{karakas_lugaro2016} where the final [Ba/Sr] $\approx -0.4$ but in that case the overall [Ba/Fe] $\approx 0.1$. Mass transfer from such an AGB companion would \textit{not} produce a star that would be flagged as being an s-process-rich candidate.

\subsection{Dynamics}
\begin{figure}
	\includegraphics[width=0.47\textwidth]{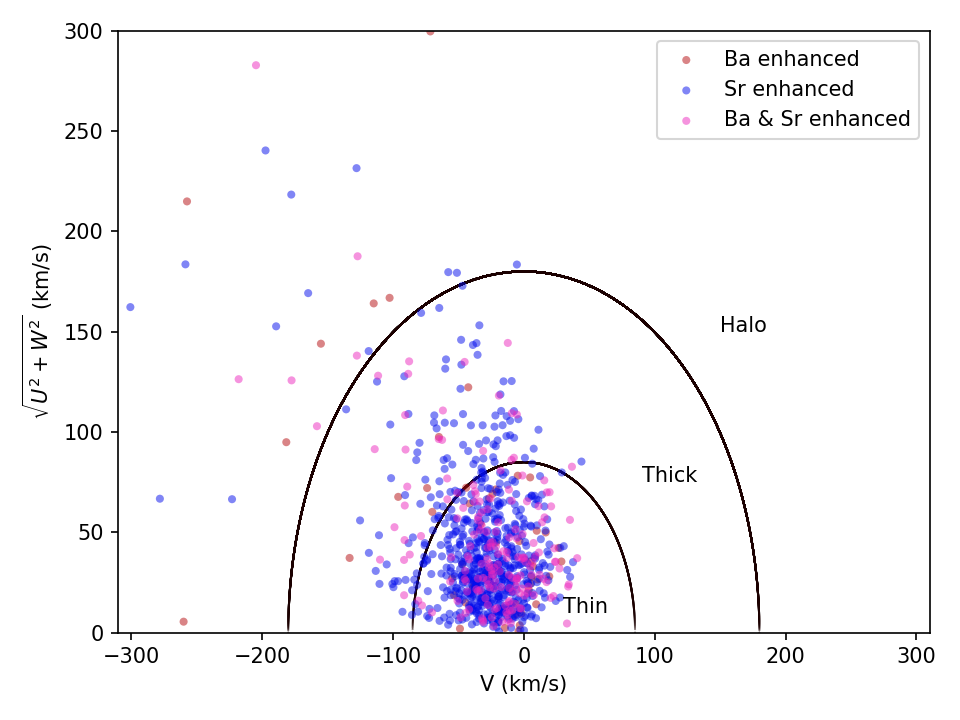}
	\caption{Galactic distribution represented by a Toomre diagram, spatial velocities (V km\,s$^{-1}$; x-axis) and ($\sqrt{U^2+W^2}$ km\,s$^{-1}$; y-axis) for number of s-process-rich candidates with Gaia DR2 parallaxes. Coloured markers red, blue, and purple highlight barium, strontium, and both barium and strontium enhancement respectively.}
	\label{fig:figure4}
\end{figure}

\begin{figure}
	\includegraphics[width=0.47\textwidth]{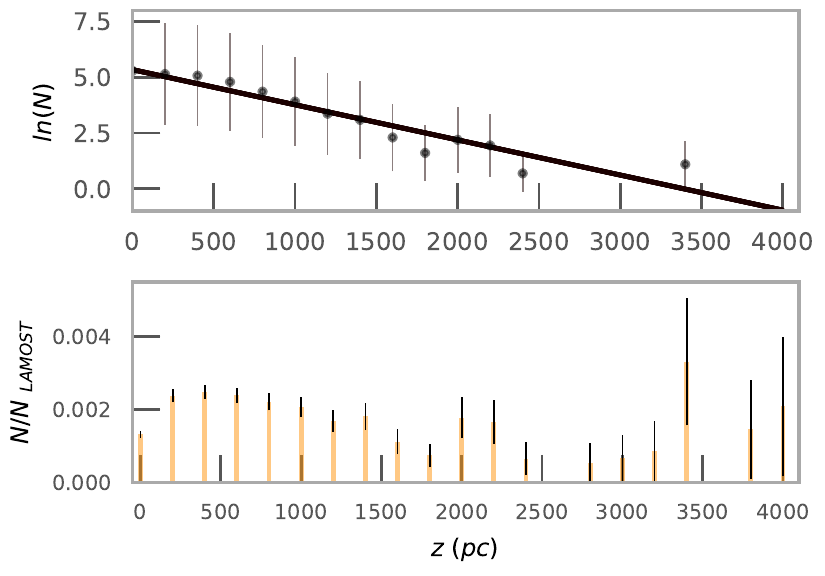}
	\caption{(a) The density distribution ($ln(N)$; y-axis) as a function of z ($|z|$; x-axis) (b) Absolute z distribution ($|z|$; x-axis) and the number of corresponding stars (normalised per galactic bin) ($N\mathbin{/}N_{LAMOST}$; y-axis). Shown for the s-process-rich candidates Gaia DR2 parallaxes, and the s-process-rich candidates as a fraction of the LAMOST sample with Gaia DR2 parallaxes.}
	\label{fig:figure5}
\end{figure}

The majority of our sample shows prograde orbits that are consistent with membership in the Milky Way disk. Figures \ref{fig:figure4} and \ref{fig:figure5} illustrate the galactic velocities for all s-process-rich candidates in this work. For our entire sample, Figure \ref{fig:figure4} shows that 97\% have velocities that are consistent with the disk, and just 3\% with velocities that are more consistent with the halo. Our sample has a similar dynamical make-up to previously published samples in the literature. \citet{gomez1997} and \citet{decastro2016} both present samples with $\sim$90\% thin disk membership, and \cite{pereira2011}'s sample of 12 s-process stars is fully consistent with disk membership. \cite{jorissen1993} attributes the galactic position of s-process enhanced stars as dependent on their intrinsic or extrinsic nature (see their Sect. 4.2). They estimate the scale height for their intrinsic stars, and find that they are concentrated towards the galactic plane whilst their extrinsic s-process stars are uniformly distributed in absolute galactic latitude. Figure \ref{fig:figure5} (top panel) shows the logarithmic distribution of the s-process-rich stars cross-matched with Gaia DR2 as a function of $|z|$. Here we derive the number density in bins of $\sim 0.2kpc$ and the error bars shown for the y-axis are Poisson errors. Following an exponential-decay profile we estimate a scale height of $z_h=0.634 \pm{0.063} kpc$ which is represented by the straight line in the figure. This is comparable to the $z_h=0.637^{0.056}_{-0.036} kpc$ result estimated by \citet{wang2018} for thick disk stars at $R=8kpc$ in the LAMOST DR3 sample. However, our 97\% disk star sample is primarily comprised of thin disk stars as seen in Figure \ref{fig:figure4}. If we remove the thick disk stars and recalculate the scale height, the scale height for only the thin disk stars is estimated to be $z_h=0.498 \pm{0.049} kpc$. This is more consistent with previous thin disk star scale height measurements (i.e. 220-450pc from \citet{bland-hawthorn2016}), and demonstrates that the thick disk stars in our sample skew the entire sample's scale height measurement substantially. Figure \ref{fig:figure5} (bottom panel) shows all absolute galactic z positions for our sample. After developing a normalised histogram (per galactic bin) and correcting for the LAMOST selection function, we show the absolute galactic z position when ignoring outlier bins where the s-process-rich candidates as a fraction of the LAMOST sample approaches unity.


\section{Conclusions} \label{sec:con}

We conducted the largest ever search for s-process enhanced stars using the LAMOST second data release. From 454,180 giant stars, we identify 895 s-process enhanced stars including; 187 s-process-rich candidates, 49 barium-enriched stars, and 659 stars with strontium enrichment. This sample size is the greatest number of s-process-rich candidates known. We estimate only a small fraction ($\sim$0.5\%) of binary configurations are favourable for s-process enriched stars. We found 97\% of our sample to have velocities that are consistent with disk membership, and just 3\% with velocities more consistent with the halo. The scale height for our entire sample is calculated to be $z_h=0.634 \pm{0.063} kpc$.

We find the majority (95\%) of our s-process candidates show carbon enhancement. Contrary to previous works we do not find s-process-rich candidates to have significantly higher [Na/Fe] than Milky Way field giants. Only 5 ($<$3\%) of our s-process-rich candidates show enhancement in Na, and the flux residuals in two of those are likely due to interstellar gas. We suggest that systematic effects in measuring [Na/Fe] from stars with low $\log{g}$ in previous works contributed to the discrepancy with our results, as well as comparisons between studies that make use of different abundance analysis techniques.

Despite our noisy estimates of [Ba/Fe] and [Sr/Fe] from LAMOST spectra, comparisons with AGB yields indicate the main neutron source responsible for s-process enhancement is consistent with the $\text{C}^{13}$(\textrm{$\alpha$},n)$\text{O}^{16}$ reaction chain. Theoretical yields suggest the progenitors of our s-process enhanced sample are low-mass AGB stars between $1 - 3\,M_{\odot}$. We encourage follow-up high resolution spectrographic observations to precisely measure a full suite of neutron-capture abundances and perform a comprehensive comparison of AGB star models.

\section*{Acknowledgements}
We thank David W. Hogg (NYU) and Hans-Walter Rix (MPIA) for useful discussions. 
A.~R.~C. is supported through an Australian Research Council Discovery Project under grant DP160100637. AIK acknowledges financial support from the Australian Research Council (DP170100521).
A. Y. Q. H. was supported by the GROWTH project funded by the National Science Foundation under PIRE Grant No 1545949, and a National Science Foundation Graduate Research Fellowship under Grant No. DGE-1144469. 
A.P.J. is supported by NASA through Hubble Fellowship grant HST-HF2-51393.001 awarded by the Space Telescope Science Institute, which is operated by the Association of Universities for Research in Astronomy, Inc., for NASA, under contract NAS5-26555.
This research has made use of NASA's Astrophysics Data System.
Guoshoujing Telescope (the Large Sky Area Multi-Object Fiber Spectroscopic Telescope LAMOST) is a National Major Scientific Project built by the Chinese Academy of Sciences. Funding for the project has been provided by the National Development and Reform Commission. LAMOST is operated and managed by the National Astronomical Observatories, Chinese Academy of Sciences. 
This work has made use of data from the European Space Agency (ESA) mission
{\it Gaia} (\url{https://www.cosmos.esa.int/gaia}), processed by the {\it Gaia}
Data Processing and Analysis Consortium (DPAC,
\url{https://www.cosmos.esa.int/web/gaia/dpac/consortium}). Funding for the DPAC
has been provided by national institutions, in particular the institutions
participating in the {\it Gaia} Multilateral Agreement.

\bibliographystyle{mnras}

\begin{thebibliography}{99}

\bibitem[\protect\citeauthoryear{Abate et al.}{2013}]{abate2013}
Abate, C., Pols, O.~R., Izzard, R.~G., Mohamed, S.~S. \& de Mink, S.~E. 2013, 
A$\&$A, 552, A26
\bibitem[\protect\citeauthoryear{Abate et al.}{2015}]{abate2015}
Abate, C., Pols, O.~R., Stancliffe, R.~J., Izzard, R.~G., Karakas, A.~I., Beers, T.~C., \& Lee, Y.~S. 2015, 
A$\&$A, 581, A62
\bibitem[Alam et al.(2015)]{alam2015} Alam, S., Albareti, F.~D., Allende Prieto, C., et al.\ 2015, \apjs, 219, 12 
\bibitem[\protect\citeauthoryear{Antipova et al.}{2004}]{antipova2004}
Antipova, L.~I., Boyarchuk, A.~A., Pakhomov, Yu.~V.,\& Panchuk, V.~E. 2004, 
ARep, 48, 597
\bibitem[\protect\citeauthoryear{Arnould et al.}{2004}]{arnould1999}
Arnould, M., Goriely, S.,\& Jorissen, A. 1999, 
A$\&$A, 347, 572
\bibitem[\protect\citeauthoryear{Badenes et al.}{2018}]{badenes2018}
Badenes, C., et al. 2018, 
ApJ, 854, 147
\bibitem[\protect\citeauthoryear{Bernstein et al.}{2003}]{bernstein2003}
Bernstein, R., Shectman, S.~A., Gunnels, S.~M., Mochnacki, S.,\& Athey, A.~E. 2003, 
SPIE, 4841, 1694    
\bibitem[\protect\citeauthoryear{Bidelman \& Keenan}{1951}]{Bidelman1951}
Bidelman, W.~P. \& Keenan, P.~C. , 1951, ApJ, 114, 473
\bibitem[\protect\citeauthoryear{Bisterzo et al.}{2014}]{bisterzo2014}
Bisterzo, S., et al. 2014, 
ApJ, 787, 10
A$\&$A, 424, 727
\bibitem[Blanco-Cuaresma et al.(2014)]{ispec} Blanco-Cuaresma, S., Soubiran, C., Heiter, U., \& Jofr{\'e}, P.\ 2014, \aap, 569, A111 
\bibitem[\protect\citeauthoryear{Bland-Hawthorn \& Gerhard}{2016}]{bland-hawthorn2016}
Bland-Hawthorn ~J.,\& Gerhard, ~G. 2016, 
\araa, 54, 529
\bibitem[\protect\citeauthoryear{Boffin \& Jorissen}{1988}]{boffin1988}
Boffin H, M.~J.,\& Jorissen, A. 1988, 
A$\&$A, 205, 155
\bibitem[\protect\citeauthoryear{B\"ohm-Vitense}{1980}]{bohm1980}
B\"ohm-Vitense, E. 1980, 
ApJ, 239, L79
\bibitem[\protect\citeauthoryear{B\"ohm-Vitense et al.}{1984}]{bohm1984}
B\"ohm-Vitense, E., Nemec, J.,\& Proffitt, C. 1984, 
ApJ, 278, 726
\bibitem[\protect\citeauthoryear{Bovy}{2015}]{bovy2015}
Bovy, J. 2015, 
ApJ, 216, 29
\bibitem[\protect\citeauthoryear{Boyarchuk et al.}{2002}]{boyarchuk2002}
Boyarchuk, A.~A., Pakhomov, Y.~V., Antipova, L.~I.,\& Boyarchuk, M.~E. 2002, 
ARep, 46, 819
\bibitem[\protect\citeauthoryear{Busso et al.}{1999}]{busso1999}
Busso, M., Gallino, R., \& Wasserburg, P.~C. 1999, 
ARA$\&$A, 37, 239
\bibitem[\protect\citeauthoryear{Busso et al.}{2001}]{busso2001}
Busso, M., Gallino, R., Lambert, D.~L., Travaglio, C.\& Smith, V.~V. 2001, 
ApJ, 557, 802
\bibitem[\protect\citeauthoryear{Casey}{2014}]{casey2014}
Casey, A.~R. 2014, 
PhD Thesis, Australian National University
\bibitem[\protect\citeauthoryear{Cristallo et al.}{2015}]{cristallo2015}
Cristallo, S., Straniero, O., Piersanti, L.,\& Gobrecht, D. 2015, 
ApJ, 219, 40
\bibitem[\protect\citeauthoryear{Cropper et al.}{2018}]{cropper2018}
Cropper, M., et al. 2018,
A$\&$A, 616, A5
\bibitem[\protect\citeauthoryear{Cseh et al.}{2018}]{cseh2018}
Cseh, B., et al. 2018,
A$\&$A, accepted
\bibitem[\protect\citeauthoryear{de Castro et al.}{2016}]{decastro2016}
de Castro, D.~B., Pereira, C.~B., Roig, F., Jilinski, E., Drake, N.~A., Chavero, C.,\& Sales Silva, J.~V. 2016, 
MNRAS, 459, 4299
\bibitem[\protect\citeauthoryear{Denissenkov \& Ivanov}{1987}]{denissenkov1987}
Denissenkov P.~A.,\& Ivanov, V.~V. 1987, 
SvAL, 13, 214
\bibitem[\protect\citeauthoryear{El Eid \& Champagne}{1995}]{el1995}
El Eid, M.~F.,\& Champagne, A.~E. 1995, 
ApJ, 451, 298
\bibitem[\protect\citeauthoryear{Fishlock et al.}{2014}]{fishlock2014}
Fishlock, C.~K., Karakas, A.~I., Lugaro, M.,\& Yong, D. 2014, 
ApJ, 797, 44
\bibitem[\protect\citeauthoryear{Gaia Collaboration et al.}{2016}]{gaia2016}
Gaia Collaboration, Prusti, T., de Bruijne, J.~H.~J., Brown, A.~G.~A., Vallenari, A., Babusiaux, C., Bailer-Jones, C.~A.~L., Bastian, U., Biermann, M.,\& Evans, D.~W. et al. 2016, 
A$\&$A, 595, A1
\bibitem[\protect\citeauthoryear{Gaia Collaboration et al.}{2018b}]{gaia2018b}
Gaia Collaboration, Brown, A.~G.~A., Vallenari, A, Prusti, T., de Bruijne, J.~H.~J., Babusiaux, C.,\& Bailer-Jones, C.~A.~L. 2018, 
\bibitem[\protect\citeauthoryear{Garc{\'{\i}}a-Hern{\'a}ndez et al.}{2013}]{garcia2013}
Garc{\'{\i}}a-Hern{\'a}ndez, et al. 2013, 
A$\&$A, 555, L3
\bibitem[Gustafsson et al.(2008)]{marcs} Gustafsson, B., Edvardsson, B., Eriksson, K., et al.\ 2008, \aap, 486, 951 
\bibitem[\protect\citeauthoryear{Gomez et al.}{1997}]{gomez1997}
Gomez, A.~E., Luri, X., Grenier, S., et al. 1997, 
A$\&$A, 319, 881
\bibitem[\protect\citeauthoryear{Han et al.}{1995}]{han1995}
Han, Z., Eggleton, P.~P., Podsiadlowski, P.,\& Tout, C.~A. 1995, 
MNRAS, 277, 1443
\bibitem[\protect\citeauthoryear{Hernquist}{1990}]{hernquist1990}
Hernquist, L. 1990, 
ApJ, 356, 359
\bibitem[\protect\citeauthoryear{Herwig}{2005}]{herwig2005}
Herwig, F. 2005, 
ARA$\&$A, 43, 435
\bibitem[\protect\citeauthoryear{Ho et al.}{2017}]{ho2017}
Ho, A.~Y.~Q., Ness, M.~K., Hogg, D.~W., Rix, H.-W. Liu, C., Yang, F., Zhang, Y., Hou, Y.,\& Wang, Y. 2017,
ApJ, 836, 5
\bibitem[\protect\citeauthoryear{Jorissen \& Boffin}{1992}]{jorissen1992}
Jorissen, A.,\& Boffin H, M.~J., 1992, 
Evidences for interaction among wide binary systems: To Ba or not to Ba? In: Duquennoy, A., Mayor, M.,(eds.) Binaries as tracers of stellar formation. Cambridge Univ. Press., p.185
\bibitem[\protect\citeauthoryear{Jorissen et al.}{1993}]{jorissen1993}
Jorissen, A., Frayer, D.~T., Johnson, H.~W., Mayor, M.,\& Smith, V.~V. 1993, 
A$\&$A, 271, 463
\bibitem[\protect\citeauthoryear{Jorissen et al.}{2005}]{jorissen2005}
Jorissen, A., Zacs, L., Udry, S., Lindgren, H.,\& Musaev, F.~A. 2005, 
A$\&$A, 441, 1135
\bibitem[\protect\citeauthoryear{Jorissen et al.}{2019}]{jorissen2019}
Jorissen, A., Boffin, H.~M.~J., Karinkuzhi, D., Van Eck, S., Escorza, A., Shetye, S.,\& Van Winckel, H. 2019, 
A$\&$A, 626, A127
\bibitem[\protect\citeauthoryear{Karakas et al.}{2000}]{karakas2000}
Karakas, A.~I., Tout, C.~A., \& Lattanzio, J.~C. 2000, 
MNRAS, 316, 689
\bibitem[\protect\citeauthoryear{Karakas et al.}{2002}]{karakas2002}
Karakas, A.~I.,\& Lattanzio C.~J.,\& Pols, O.~R. 2002, 
PASA, 515, 19
\bibitem[\protect\citeauthoryear{Karakas \& Lattanzio}{2003}]{karakas2003}
Karakas, A.~I.,\& Lattanzio C.~J. 2002, 
PASA, 20, 279
\bibitem[\protect\citeauthoryear{Karakas \& Lattanzio}{2014}]{karakas2014}
Karakas, A.~I.,\& Lattanzio C.~J. 2014, 
PASA, 31, 30
\bibitem[\protect\citeauthoryear{Karakas \& Lugaro}{2016}]{karakas_lugaro2016}
Karakas, A.~I.,\& Lugaro M. 2016, 
ApJS, 825, 26
\bibitem[\protect\citeauthoryear{Karakas et al.}{2016}]{karakas2016cp}
Karakas, A.~I. 2016, 
S.A.Lt, 229, 87
\bibitem[\protect\citeauthoryear{Karakas et al.}{2018}]{karakas2018}
Karakas, A.~I., Lugaro, M., Carlos, M., Cseh, B., Kamath, D.,\& Garc{\'i}a-Hern{\'a}ndez, D.~A.  2018, 
MNRAS, 447, 421
\bibitem[\protect\citeauthoryear{Karinkuzhi et al.}{2018}]{karinkuzhi2018}
Karinkuzhi, D., et al. 2018, 
A$\&$A, 618, A32
\bibitem[\protect\citeauthoryear{Katz et al.}{2018}]{katz2018}
Katz, D., et al. 2018,
A$\&$A, 616, A11
\bibitem[\protect\citeauthoryear{Kelson et al.}{2000}]{kelson2000}
Kelson, D.~D., Illingworth, G.~D., van Dokkum, P.~G.,\& Franx, M. 2000, ApJ, 531, 159
\bibitem[Kupka et al.(1999)]{vald} Kupka, F., Piskunov, N., Ryabchikova, T.~A., Stempels, H.~C., \& Weiss, W.~W.\ 1999, \aaps, 138, 119 
\bibitem[\protect\citeauthoryear{Lindegren et al.}{2018}]{lindegren2018}
Lindegren, L., et al. 2018,
A$\&$A, 616, A2
\bibitem[\protect\citeauthoryear{Liu et al.}{2018}]{liu2000}
Liu, J.~H., Zhang, B., Liang, Y.~C., \& Peng, Q.~H. 2000,
A$\&$A, 363, 660
\bibitem[\protect\citeauthoryear{Luck \& Heiter}{2007}]{luck2007}
Luck R.~E.,\& Heiter, U. 2007, 
AJ, 133, 2464
\bibitem[\protect\citeauthoryear{Lugaro et al.}{2012}]{lugaro2012}
Lugaro, M., Karakas, A.~I., Stancliffe, R.~J., Rijs, C. 2012, 
ApJ, 747, 2
\bibitem[\protect\citeauthoryear{Luo et al.}{2015}]{luo2015}
Luo, A.~L., Bai, Z.~R., et al. 2015, 
RAA, in press
\bibitem[\protect\citeauthoryear{MacConnell et al.}{1972}]{macconnell1972}
MacConnell, D.~J., Frye, R.~L.,\& Upgren, A.~R. 1972, 
\aj, 77, 384
\bibitem[Maiorca et al.(2011)]{maiorca2011} Maiorca, E., Randich, S., Busso, M., Magrini, L., \& Palmerini, S.\ 2011, \apj, 736, 120 
\bibitem[\protect\citeauthoryear{Malaney \& Lambert}{1988}]{malaney1988}
Malaney, R.~A.,\& Lambert, D.~L. 1988, 
MNRAS, 235, 695
\bibitem[\protect\citeauthoryear{McClure}{1983}]{mcclure1983}
McClure, R.~D. 1983, 
ApJ, 268, 264
\bibitem[\protect\citeauthoryear{Mennessier et al.}{1997}]{mennessier1997}
Mennessier, M.~O., Luri, X., Figueras, F., et al. 1997, 
A$\&$A, 326, 722
\bibitem[\protect\citeauthoryear{Mishenina et al.}{2006}]{mishenina2006}
Mishenina, T.~V., Bienaym\' e, O., Gorbaneva, T.~I.,\& Charbonnel, C. 2006, 
A$\&$A, 456, 1109
\bibitem[\protect\citeauthoryear{Miyamoto-Nagai}{1975}]{miyamoto1975}
Miyamoto, M,\& Nagai, R. 1975, 
PASJ, 27, 533
\bibitem[\protect\citeauthoryear{Mowlavi}{1999}]{mowlavi1999}
Mowlavi, N. 1999, 
A$\&$A, 350, 73
\bibitem[\protect\citeauthoryear{Navarro, Frenk \& White}{1997}]{nfw1997}
Navarro, J.~F., Frenk, C.~S.,\& White, S.~D.~M. 1997, 
ApJ, 490, 493
\bibitem[\protect\citeauthoryear{Ness et al.}{2016}]{ness2016}
Ness M., Hogg D.~W., Rix H.~W., Martig M., Pinsonneault M.~H., Ho A. 2016, 
ApJ, 823, 114
\bibitem[\protect\citeauthoryear{Pereira et al.}{2011}]{pereira2011}
Pereira, C.~B., Sales Silva, J.,~A., Chavero, C., Roig, F.,\& Jilinski E. 2011, 
A$\&$A, 533, A51
\bibitem[\protect\citeauthoryear{Pourbaix et al.}{2004}]{pourbaix2004}
Pourbaix, D., Tokovinin, A.~A., Batten, A.~H., Fekel, F.~C., Hartkopf, W.~I. et al. 2001, 
\bibitem[\protect\citeauthoryear{Price-Whelan}{2014}]{mwpotential2014}
Price-Whelan, A.~M., Hogg, D.~W., Johnston, K.~V.,\& Hendel, D. 2014, 
Apj, 4, 794
\bibitem[\protect\citeauthoryear{Price-Whelan}{2017}]{price2017}
Price-Whelan, A.~M. 2017, 
The Journal of Open Source Software, 2, 388
\bibitem[\protect\citeauthoryear{Sartoretti et al.}{2018}]{sartoretti2018}
Sartoretti, P., et al. 2018,
A$\&$A, 616, A6
\bibitem[\protect\citeauthoryear{Schectman \& Johns}{2003}]{schectman2003}
Shectman, S.~A.,\& Johns, M. 2003, 
SPIE, 4837, 910
\bibitem[\protect\citeauthoryear{Schlafly \& Finkbeiner}{2011}]{schlafly2011}
Schlafly, E.~F.,\& Finkbeiner, D.~P. 2011, 
Apj, 737, 103
\bibitem[\protect\citeauthoryear{Smiljanic}{2012}]{smiljanic2012}
Smiljanic, R. 2012, 
MNRAS, 422, 1562
\bibitem[\protect\citeauthoryear{Sneden et al.}{2008}]{sneden2008}
Sneden, C., Cowan, J.~J., \& Gallino, R.\ 2008,
\araa, 46, 241
\bibitem[\protect\citeauthoryear{Slemer et al.}{2017}]{slemer2017}
Slemer, A., et al. 2017, 
MNRAS, 465, 4817
\bibitem[\protect\citeauthoryear{Straniero et al.}{2014}]{straniero2014}
Straniero, O., Cristallo, S. \& Piersanti, L. 2014, 
ApJ, 785, 77
\bibitem[\protect\citeauthoryear{Takeda et al.}{2008}]{takeda2008}
Takeda, Y., Sato, B.~V.,\& Murata, D. 2008, 
AJ, 60, 781
\bibitem[\protect\citeauthoryear{Tian et al.}{2018}]{tian2018}
Tian, Z., et al. 2018, 
RAA, 18, 52
\bibitem[\protect\citeauthoryear{Travaglio et al.}{2001}]{travaglio2001}
Travaglio, C., et al. 2001, 
ApJ, 549, 346
\bibitem[\protect\citeauthoryear{Wang et al.}{2018}]{wang2018}
Wang, H-F., Liu, C., Xu, Y., Wan, J-C.,\& Deng, L. 2018 
MNRAS, 478, 3367
\bibitem[\protect\citeauthoryear{Webbink}{1986}]{webbink1986}
Webbink, R.~F. 1986, 
In: Leung, K.~C., Zhai, D.~S.(eds.) Critical Observations versus Physical Models for Close Binary Systems. Gordon and Breach, New York, p.403
\bibitem[\protect\citeauthoryear{Whitelock et al.}{2013}]{whitelock2013}
Whitelock, P.~A., et al. 2013, 
MNRAS, 428, 2216
\bibitem[\protect\citeauthoryear{Wu et al.}{2011a}]{wu2011a}
Wu, Y, et al. 2011, 
Research in Astronomy and Astrophysics, 11, 924
\bibitem[\protect\citeauthoryear{Wu et al.}{2011b}]{wu2011b}
Wu, Y, Singh, H.~P., Prugniel, P., Gupta, R.,\& Koleva, M. 2011, 
A$\&$A, 525, A71
\bibitem[Valenti \& Piskunov(1996)]{sme} Valenti, J.~A., \& Piskunov, N.\ 1996, \aaps, 118, 595 
\bibitem[\protect\citeauthoryear{Van Eck \& Jorissen}{1999}]{van1999}
Van Eck, S. \& Jorissen, A. 1999, 
A$\&$A, 345, 127
\bibitem[\protect\citeauthoryear{Van der Swaelmen et al.}{2017}]{van2017}
Van der Swaelmen, M., Boffin, H.~M.~J., Jorissen, A., \& Van Eck, S. 2017
A$\&$A, 597, A68
\bibitem[\protect\citeauthoryear{Ventura et al.}{2013}]{ventura2013}
Ventura, P., Di Criscienzo, M., Carini, R.,\& D'Antona, F. 2013, 
MNRAS, 431, 3642

\end{thebibliography}

\bsp	
\label{lastpage}
\end{document}